\documentclass[intlimits,twoside,a4paper]{article}
\usepackage{amsmath,amssymb}
\usepackage{graphicx}

\usepackage[T2A]{fontenc}
\usepackage[cp1251]{inputenc}

\usepackage{cmpj2}

\issue{2015}{18}{1}{13001}
\doinumber{10.5488/CMP.18.13001}

\title[Gone but not forgotten]{Gone but not forgotten}

\author[D. Henderson]{D. Henderson\thanks{E-mail: doug@chem.byu.edu}}
\address{Department of Chemistry and Biochemistry,
Brigham Young University, Provo UT 84602-5700}

\date{Received June 16, 2014}
\authorcopyright{D. Henderson, 2015}

\begin{document}

\maketitle


\begin{abstract}

In this reminiscence I discuss the influence of Henry Eyring and
John Barker upon my life and work.  Others, especially my
family, have been of even
greater personal influence.  However, these two great and grand men were of
tremendous scientific influence.  Of course, others who came before Eyring and
Barker, especially Boltzmann and van der Waals and later Onsager and
Eyring's contemporary,
Kirkwood, have been influential, but only indirectly
as I never met them.  Eyring and Barker are not the only scientists who have
inspired me.  Many who influenced me have contributed articles to this
special issue or have worked with me.  I single out Eyring and Barker because
I met them early in my career and because they have passed away and are now
present only in spirit.  They are gone but should not be forgotten; I take
this occasion to remind the readers about these two outstanding scientists
and fine men and offer this reminiscence as thanks to them.


\pacs 01.60.+q, 01.65.+g, 64.10.+h, 82.20.-w, 82.20.Kh

\end{abstract}

Isaac Newton, in one of his few modest moments, said that the reason that he
could see so far is because he stood on the shoulders of those who preceeded
him.  This is true of us all.  Boltzmann and van der Waals have provided
me with great insight.  More recently, Onsager and Kirkwood, provided
foundations
upon which I have built.  However, I know those gentlemen only through
their work.  I knew and worked with both Eyring and Barker and admired them
greatly.  Of
course, I have worked with and admired many others.  Many have contributed
to this issue.  Henry Eyring and John Barker were important
scientists in my career and good friends but have passed away and cannot
participate in this volume.  They have gone but are not forgotten.

\section*{Henry Eyring (1901--81)}

\begin{figure}[!t]
\begin{center}
\includegraphics[width=7.6cm,clip]{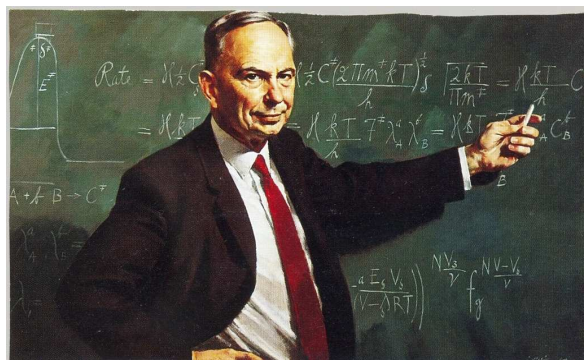}
\caption{Henry Eyring.}\label{fig1}
\end{center}
\end{figure}

Henry Eyring appears in figure~\ref{fig1}.  This is a painting that hangs in
foyer of the Universiy of Utah chemistry building.

I first became aware of Henry Eyring when I was quite young.  My mother told
me excitedly
that she had learned in a church class that one could prove the existence
of God by science.  Prove is really too strong a word.  My mother
misunderstood.  Rather the author of
the lesson used the example of a power series to demonstrate that by one
could approach God in successive, hopefully convergent, steps.  Later, when
talking with Henry Eyring, I learned that he was the author of this lesson.

My first real contact with Henry was as an undergraduate at the University of
British Columbia, when Frank Peters, a friend who
was majoring in metallurgy, drew my attention to the fact that Henry Eyring,
a member of my church, was to be a seminar speaker at the university.  Frank
was impressed by Henry's breadth of knowledge, his lucid presentation of his
ideas, and his friendly manner.   It was no coincidence that a few years
later I found myself to be Eyring's research student at the University of Utah.

Eyring was not born into a scientific family.  He was born in Colonia Juarez,
a small town
in northern Mexico.  Colonia Juarez may be found by looking at a map of
northern Mexico and following the railway line southwest from El Paso for
around   200 kilometers to Nuevos Casas Grandes and then going west on a
secondary road for about 20 kilometers.  Colonia Juarez is at the end of
this road.  The major industry of the region is the growing of fruit.
However, Henry's father was a rancher with a large herd of cattle.
The railroad was very valuable for the export of produce from the region.

At the time of Henry's birth, and to a somewhat lesser extent today, Colonia
Juarez was populated mostly by Americans and decendents of Americans.  The
word ``Colonia'' gives the impression to English speakers of a colony of
people who came to exploit the locals, get rich and then retire to
their homeland.  People in Utah tend to refer to Colonia Juarez and
similar towns as the ``Mexican colonies''.  However, in Mexico, the word
colonia has the connotation of a neighborhood.
For example, the neighborhood of Mexico City in which Rose-Marie and I lived
for a few years is Colonia del Valle.
The residents of Colonia Juarez came to stay.   Henry's father became a Mexican
citizen.  Henry was a Mexican citizen until he became a naturalized US citizen
in the mid 1930's.  The
Mexican revolution changed things and, for safety, the Eyring family departed
hurriedly  on the railroad with their friends and neighbors and went to El
Paso.  Many of his relatives returned to Mexico when the
situation stabilized but after some soul searching Henry's family went to
Arizona, where they cleared their land, farmed and lived in reduced
circumstances.  It is worth noting that Henry contributed to the payment of
the mortgage on the farm for a number of years after he left home.

Henry went first to the University of Arizona in Tucson and then to the
University of California in Berkeley, where he obtained his PhD in chemistry.
A few years after graduation, he married Mildred Bennion; they went to
Germany, where he was a post-doctoral fellow
at the Kaiser Wilhelm Institute (now the Max Planck Institute) in Dahlem, a
suburb of Berlin.
While there, he and Polanyi \cite{1} made the first quantum mechanical calculation
of the energy surface for a chemical reaction.  On his return to the US,
he spent a short stay  at Berkeley, where his first son was born, and then
settled in as a faculty member at Princeton, where he ultimately became a
professor of chemistry.  While at Princeton his family grew with the
addition of two more sons.  At Princeton, he developed \cite{2} what he called
absolute reaction rate theory, with which his name is forever associated.
Either the Eyring-Polanyi energy surface calculation or his rate theory was
worthy of a Nobel Prize but one never came.  Some of Henry's associates
were shocked (their word) by this oversight.  Henry accepted this in good
grace.

By the time that I met Henry, he had moved to the University of Utah, where
he became the founding Dean of the Graduate School of the university.
Although he was a dean,
he still taught classes in his office and had PhD students.  I met Henry when
he wore all these multiple hats.  While at Princeton, Henry became interested
in developing a theory of the liquid state.  This interest was due to the
fact that his reaction rate theory required the partition functions of
the reactants.  His classic rate theory paper considered low density gases
and needed
only the partition function of an ideal gas.  However, many important reactions
occur in liquids and for this the partition function of a liquid is needed.
At Princeton he developed a cell model of a liquid.  When applied to a low
density fluid, this model lacks a factor, $\exp(N)$, that  appears
in the entropy of an ideal gas.  This factor should appear in
a natural way but Eyring reasoned that inserting the factor, even arbitrarily,
was preferable to not having it at all.  He called this rather arbitrary term
the {\it communal entropy} because it was a measure of the multiple occupancy
of the cells and was shared by all the molecules.

Subsequently, Eyring thought that a lattice model should have empty sites or
cells with holes.  He reasoned that, when a liquid molecule evaporated, it left
an empty cell behind.  He regarded these holes as approximate mirror images of
the vapor molecules.  The entropy of mixing of the molecules and holes
provided the
communal entropy.  If the liquid consisted of molecules and holes that
{\it exactly} mirrored the vapor molecules, the sum of the densities of the
coexisting liquid and vapor phases
should be a constant.  This provided a simple qualitative explanation of the
law of rectilinear diameters, where the observed sum of the densities of the
coexisting
vapor and liquid is not a constant but is a linear function of the
temperature.  Extending this idea, Eyring suggested that the partition
function, $Z$,  of a liquid could be written as

\begin{equation}
Z=Z_\textrm{s}^{\frac{V_\textrm{s}}{V}}Z_\textrm{g}^{\frac{V-V_\textrm{s}}{V}},
\label{eq:1}
\end{equation}
where $Z_\textrm{s}$, $Z_\textrm{g}$, $V_\textrm{s}$ and $V$ are, respectively, the partition functions
 of the solid and ideal gas
phases, and the molar volumes of solid and liquid phases.  For $Z_\textrm{s}$ and $Z_\textrm{g}$,
the partition functions of the Einstein model of a solid and the ideal
gas were used.  Empirical parameters were used in $Z_\textrm{s}$ and for $V_\textrm{s}$.

Equation (\ref{eq:1}) suggests that the heat capacity, $C$, of liquid  argon should be

\begin{equation}
\frac{C}{Nk}=3\frac{V_\textrm{s}}{V}+\frac{3}{2}\frac{V-V_\textrm{s}}{V},
\label{eq:2}
\end{equation}
where $N$ is the number of molecules in the liquid, $k$ is Boltzmann's
constant, and $T$ is the temperature of the liquid.  This result is remarkedly
accurate.  It does fail to predict the singularity
of the heat capacity at the critical point.  Eyring's critics made too much
of this `failing' since their theories also shared this failing.  Eyring called
this procedure the {\it significant structure theory} of the liquid state \cite{3}.
The name was chosen because Eyring thought that the molecules and holes
were the significant structural elements of a liquid.
I was put to work applying these ideas and presented the result for a thesis.
The prospect interested me because at the time the lack of a theory of the
liquid state
was regarded as one of the major stumbling blocks in science.  For me,
significant structure theory did not remove this stumbling block.
The removal of the stumbling block
came when I worked with Barker a few years later and when he and I had
spectacular (John Rowlinson's term) success.

Significant structure theory is not a true theory; it is a description.
It provides
no insight into the relation of the thermodynamic properties of a liquid to
the intermolecular interactions between the liquid molecules.  It is an
interpolation formula to pass between the solid and gas phases.  The
critics of significant structure theory were too harsh because it does
focus on the essential importance of the volume or density and assigns a
lesser role to the temperature in determining the properties of
a liquid.  This became clear as a result of the work of Barker and me.  At
the time of my graduate work, the reverse was thought to be the case.
Also, it is worth
noting that equation (\ref{eq:1}) implies that the portion of the free energy of a liquid
that results from those molecules that have not evaporated
contains a term that is proportional to $V_\textrm{s}/V$, or to the density, of the
liquid.  This is similar to the ideas of van der Waals.  I will return
to this point shortly.

After graduation, I tried to develop a more rigorous theory of the liquid
state that involved holes or vacant cells.  I had mixed success in this
endevour but I did publish a paper that captured the interest of John Barker.
Despite my apostasy from significant structure theory, Henry and I remained
close.  He invited me to join with him and Jost in editing a multi-volume
treatise on physical chemistry.  I was nervous about this and wondered if
I was into something that was over my head.  However, it worked out.  The
treatise was a success.

Despite the fact that, as time passed, Henry's work had become somewhat out
of the main stream of theoretical chemistry, I really enjoyed my association
with him.  He was an exceptionally kind man.  He helped me and my wife when we
needed his support.  He was modest and did not feel that he was any
better than anyone else.  He was genuinely interested in the personal and
professional details of anyone he met and spoke with.  He had a facility to
make a person feel at ease in his presence.

He was also an engaging speaker.  Perhaps he was at his best when he spoke
about his research at Princeton into the effect of pressure on the behavior of
tadpoles.  The tadpoles would swim normally at atmospheric pressure but
under pressure they behaved erratically and then behaved normally again when
the pressure was reduced.  Henry, when speaking about this experiment would act
the part of the tadpoles and alternate between a normal state and an apparently
intoxicated state while commenting on the effect of alcohol on the brain (he
did not use or approve of alcohol as a beverage).  The talk was hilarious and
it was obvious that he was enjoying himself enormously.

\section*{John Adair Barker (1925--95)}

\begin{figure}[htb]
\begin{center}
\includegraphics[width=6.0cm,clip]{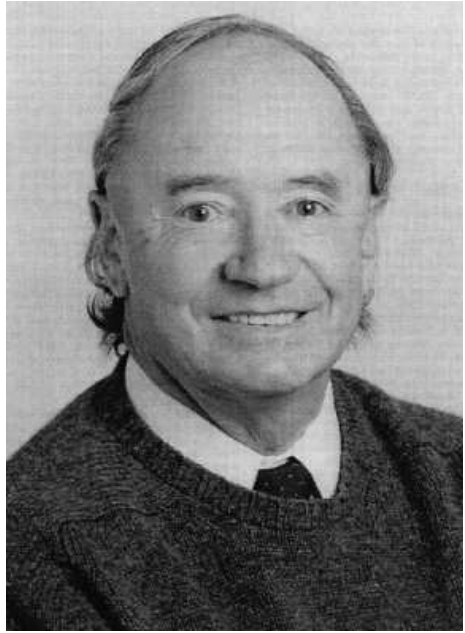}
\caption{John Barker.}\label{fig2}
\end{center}
\end{figure}

John Barker appears in figure~\ref{fig2}, a photograph that was given to me by
his widow, Sally.

John was born in Corrigin, Western Australia, which is
about 200 kilometers southeast of Perth.  While he was young, he and his
family moved to Red Cliffs in Victoria, Australia.  Red Cliffs is near
Mildura and is about 500 kilometers northwest of Melbourne.  Similarly to
Eyring, he grew up in a relatively isolated farming community.  In fact,
prior to jet
aircrafts, Australia was perhaps even more isolated than Colonia Juarez.   His
father was a country medical doctor.  John attended high school at a boarding
school in Melbourne and then attended the
University of Melbourne.  At least once, presumably as an adventure, he
chose to return home to Red Cliffs for school vacations by bicycle.  His
friends in Australia are still in awe of this feat that is indicative of John's
vigor.

John graduated from the University of Melbourne in physics and mathematics.
One of his professors was H.H.~Corben, the author of an influential book on
classical mechanics.  Interestingly, I met Bert Corben and his wife by chance
some years later in a restaurant in Mexico.
John might have proceeded immediately to a doctoral degree.  However, a family
tragedy led to his spending a, not very happy, year teaching mathematics at a
secondary school in England.  During the long trip, by ship, back to Australia,
he met Sally Johnston, whom he married soon
after they had both arrived in Australia.  They had two sons and a daughter.
Since he now had a family to support, graduate studies were out of the
question so he joined CSIR (now CSIRO), a federal
government scientific organization, as a scientific officer in Melbourne.
He was initially hired to provide theoretical support to the experimental
work of Ian Brown.  However, John was soon working independently and rose
through the ranks of CSIRO to become a chief research officer.  Several years
after joining CSIRO, he
presented a summary of his research as a thesis and was awarded a DSc degree
by the University of Melbourne.

Some of his early work at CSIRO involved the use of perturbation expansions.
However, like Eyring and many other contemporaries, Barker felt that since the
density of a liquid was similar to that of a solid, a lattice approach
was the appropriate starting point for the development of a theory of the
liquid state.  His approach was more mathematical than the intuitive style
of Eyring.  Barker started with a lattice but attempted to introduce disorder
and multiple occupancy in a systemmatic manner.  Although perturbation
expansions were employed in his lattice theory of liquids, Barker did not
fully return to perturbation theory until I appeared on the scene about
fifteen years later.

As I have mentioned, after obtaining my doctorate, I attempted to develop,
with mixed success, the ideas of Eyring
regarding holes in a lattice in a more systematic manner.  Soon after
graduation I published a paper that caught Barker's attention.  He, and
the Chief of the Physical Chemistry Division of CSIRO, Sefton Hamann,
wrote to me and invited me to come to Melbourne and work with Barker.  I
replied positively and enthusiastically since I had admired the work of
Hamann and Barker for
some time but I was married and had two (soon three) small children and
could not just take up and go.  However, I arranged my affairs and as soon
as I had accumulated enough `credits' at the University of Waterloo for a
sabbatical, and with the assistance of Alfred P.~Sloan Foundation and Ian
Potter Foundation fellowships, I went to Melbourne to work with John.

While at CSIRO in Melbourne, I was asked to give a series of lectures.
By this time, I was drifting away from lattice theories.  In my lectures, I
discussed some recent work on perturbation theory about which I had just
learned prior to coming to Melbourne.  I discussed some recent papers of
Rowlinson \cite{4}, Frisch {et al.} \cite{5}, and McQuarrie and Katz \cite{6} who had
attempted to start with a hard sphere fluid as an unperturbed or reference
system and apply a
perturbation expansion.  The hard sphere fluid was an attractive starting
point because there were accumulating data from simulations for the
thermodynamics and structure of this fluid and, especially through the work of
Wertheim \cite{7,8}, convenient and accurate theoretical expressions for their
thermodynamics and distribution functions.  At the time the
evidence was that these perturbation theories were interesting but applicable
only to gas well above its critical temperature but
not to a liquid because the temperature of a liquid was apparently too low for a
perturbation
series to converge.  This led to Barker and me wondering whether this was
due to the manner in which the repulsive part of intermolecular potential
was treated and to the manner in which the intermolecular potential was divided
into the potential of the unperturbed or reference fluid and the
perturbation energy.

We decided
that, to avoid these questions, we should consider first a system in which
the repulsion was exactly hard.  Before I went to Melbourne, Berni Alder had
told me of
his (as yet unpublished) molecular dynamics simulations for a hard
sphere fluid with a square well attractive potential.  Such a square well
fluid seemed ideal for our needs.  I wrote Berni and he
generously agreed to provide his
results in advance of publication.   We found that the agreement of our
expansion, to first order, with the, as yet unpublished, simulation results
 of Alder and the earlier simulations of
Rotenberg \cite{9} was promising and, with the inclusion of
an intuitively attractive approximation for the second order term,
the results of our version of perturbation theory gave excellent results
for the entire liquid region \cite{10}.

Thus encouraged, Barker and I devised a scheme for treating the
fact of the repulsive part of a more realistic intermolecular potential.  The
repulsive part of the intermolecular interaction of a realistic fluid,
although very steep, is not infinitely steep.  In our scheme, the
perturbation expansion was formally the same as the earlier expansion
based on a hard sphere reference fluid but now the diameter of the unperturbed
hard sphere fluid was a function of the temperature.

We obtained \cite{11} the following expression for the Helmholtz function of a fluid.

\begin{equation}
\frac{A-A_0}{NkT}=\frac{1}{2}\rho\beta\int u_1(R)g_0(R)\rd{\bf R}
-\frac{1}{4}\rho\beta^2\int [u_1(R)]^2 kT
\left[\frac{\partial \rho g(R)}{\partial p}\right]_0\rd{\bf R}+\cdots,
\label{eq:3}
\end{equation}
where $\rho=N/V$, the subscript 0 indicates that the quantity is that of
the reference hard sphere fluid, and ${\bf R}$ is the distance between the
centers of the molecules (or hard spheres).  The function $g_0(R)$ is
the radial distribution function of the reference fluid.  The radial
distribution function is strongly dependent on the density.  However, the
integral of $u_1(R)g_0(R)$ is nearly constant.   The approximate second order
term, given in equation (\ref{eq:3}), is very useful for many systems.  We called this
approximation
the {\it compressibility approximation}.  This approximation can
be made even more simple by neglecting the density dependence of $g_0(R)$ in the
the integral in the
second order term.  Praestgaard and Toxvaerd \cite{12} extended the compressibility
approximation to obtain an estimate for the higher order terms.

It is worth noting that if the series is truncated at first order and if
the density dependence of $g_0(R)$ in the integral is neglected, the first order
term is a linear function of $\rho$, just as it is in the van der Waals and
significant structure theories.

Perturbation theory has been very useful and important.  It was, as stated
by Rowlinson, a spectacular advance.  It was the first successful theory for
the liquid state.  My stay in Melbourne was an exciting year.  Interestingly,
our theory uncovered an error \cite{13} in
some simulations of liquid mixtures.  This is one of the very few instances that
theory has corrected a simulation.

Additionally, perturbation theory clarified the problem with integral
equation theories of liquids.  The integral equation theories tended to
yield poor results for the pressure.  The reason for this is that the
route from the radial distribution function (RDF) to the pressure is very
sensitive to errors in the RDF.  By contrast, perturbation theory relies
on the free energy, which is not so sensitive.  In fact, we \cite{14}
showed that if an integral equation approach is used to calculate the
energy and if the free energy is then obtained by numerical integration and
then the pressure is obtained by numerical differentiation, excellent
results are obtained.  Of course, all this numerical integration and
differentiation is cumbersome.  Much of the reason for the success of the
mean spherical approximation is that, for some systems, the
free energy and pressure can be obtained from the energy without numerical
integration and differentiation.

Subsequently, Weeks {et al.} \cite{15} and my former graduate student,
Bill Smith, together with Keith Gubbins and their colleagues, \cite{16} developed an
alternative version of
perturbation theory that is based on a different division of the potential into
reference and perturbation terms. This approach leads to a more rapid
convergence.  However, with the compressibility approximation for the
second order term, equation (\ref{eq:3}) is equally accurate and as user friendly as
this approach.   The two approaches are equivalent for
fluids with a hard core.

The last part of Barker's career was devoted to obtaining accurate results
for interactions between molecules and surfaces.  His expression for
the interaction potential between the inert gas molecules was highly accurate.  He
found that the evidence about the pair interactions between argon molecules
that was obtained from viscosity and scattering experiments was inconsistent.
He made a bold step of abandoning the viscosity results and proposed an
useful expression for the argon pair potential.  It is bold for a theoretician
to use theory to reject experimental results.
New viscosity experiments supported
his view.  Barker's last project concerned the interaction of gas molecules with
surfaces.  His passing prevented the completion of this work.

No matter how well you know someone, you never know everything.  After John
passed away, I learned that he wrote poetry.

\section*{Concluding remarks}

It is interesting for me to note that I am a coauthor of Eyring's last
scientific work, a book, and that Barker's last publication was a generous
appraisal of me and my work.  I have also mentioned
Sefton Hamann, another distinguished scientist.  He helped facilitate my
year in Melbourne.  I never worked closely with him but did like him.  Years
later, and after both Eyring and Barker had passed away, I had an occasion to
spend a couple of weeks  at the University of Melbourne and during my stay had
lunch with Sefton
and his wife, Elizabeth.  Sefton and I discussed how the ``established''
theories of liquid mixtures failed to account for some of the properties
of liquid mixtures that interested him.  Subsequently, Sefton and I showed
in a simple way how perturbation theory provided the insight that he was
seeking.  After Sefton's passing, I was told that this was Sefton's last
paper.  I seem to specialize in last papers.  As an added evidence of this, I
mention that I am a coauthor of the last papers of Peter Leonard, one of my
students.  My colleagues should be aware
that any paper they publish with me may be their last.  I debated about
whether I should
include Sefton more fully in this reminiscense but decided not to do so.  I
admired Sefton but did not work closely with him as I did with Henry and John.

There are some echos of Eyring in my life.  We are members of the same,
small but growing, Church.  We both lived in Mexico for a period.  He and I
worked in a copper mine while we were students.  He worked and my parents
(but not me) lived in Dahlem in west Berlin.

A comparison of Eyring and Barker is difficult and not very meaningful.
However, an attempt is required.  Eyring's work was uneven.
His later work was not as important as his early work.  Barker's work
was more even but not as influential as Eyring's early work.  In any case,
both were outstanding scientists and splendid persons and I have great
affection for both of them.

The families of Henry Eyring and John Barker have told
me that both Henry Eyring and John Barker regarded me as an honorary family
member.  I am grateful for this.  In any case, they both were admirable men
and scientists.  I am better as a scientist and a person for having known and
worked with them.


\ukrainianpart

\title{Вони відійшли, але їх не забули}

\author{Д. Гендерсон}
\address{Відділ хімії та біохімії,
 Університет Брігема Янга, Прово, штат Юта 84602-5700, США
}

\makeukrtitle

\begin{abstract}
 У цих спогадах я обговорюю той вплив, який Генрі Ейринг і Джон Баркер мали на моє життя та роботу. Хоча інші, особливо моя сім'я,
 мали навіть більший особистий вплив на мою особистість. Безперечно й інші, які прийшли ще раніше, ніж Ейринг і Баркер, особливо Больцман і Ван дер Ваальс, а пізніше Онсагер та сучасник Ейринга Кірквуд мали вплив на мене, але лише опосередковано, оскільки я ніколи не зустрічався з ними.
 Однак Ейринг та Баркер не єдині вчені, які надихали мене. Багато з тих, хто впливали на мене вже подали свої статті у цей спеціальний випуск або ж працювали зі мною. Я виокремлюю саме Ейринга і Баркера тому, що зустрівся з ними на початку моєї кар'єри і тому, що вони вже відійшли і зараз присутні лише своїм духом. Вони відійшли, але не повинні бути забутими. Я користаюсь цією нагодою, щоб нагадати читачам про цих двох визначних науковців та чудових людей і пропоную вам ці спогади, як подяку їм.

\keywords хімічна кінетика, теорія рідин, теорія збурень


\end{abstract}

\end{document}